# Direct visualization of quasiparticle concentration around superconducting vortices

*Jian-Feng Ge,[1,2] Koen M. Bastiaans, [1,3] Jiasen Niu,[1,4] Tjerk Benschop,[1] Maialen Ortego Larrazabal,[5] Milan P. Allan[1,4]\**

[1] *Leiden Institute of Physics, Leiden University, 2333 CA Leiden, The Netherlands*

[2] *Max Planck Institute for Chemical Physics of Solids, 01187 Dresden, Germany*

[3] *Department of Quantum Nanoscience, Kavli Institute of Nanoscience, Delft University of Technology, 2628 CJ Delft, The Netherlands*

[4] *Faculty of Physics, Ludwig-Maximilians-University Munich, Munich 80799, Germany*

[5] *Debye Institute for Nanomaterials Science, Utrecht University, 3508 TA Utrecht, The Netherlands*

\*Corresponding author. Email: allan@physics.leidenuniv.nl

**Bogoliubov quasiparticles play a crucial role in understanding the behavior of a superconductor, and in achieving reliable operations of superconducting quantum circuits. Diagnosis of quasiparticle poisoning at the nanoscale provides invaluable benefits in designing superconducting qubits. Here, we use scanning tunneling noise microscopy to locally quantify quasiparticles by measuring the effective charge. Using the vortex lattice as a model system, we directly visualize the spatial variation of the quasiparticle concentration around superconducting vortices, which can be described within the Ginzburg-Landau framework. This shows a direct, noninvasive approach for the atomic-scale detection of relative quasiparticle concentration as small as $10^{-4}$ in various superconducting qubit systems. Our results alert of a quick increase in quasiparticle concentration with decreasing intervortex distance in vortex-based Majorana qubits.**

Superconducting qubit states are protected from decaying by the energy gap, but pair-breaking excitations, known as Bogoliubov quasiparticles, always exist when the real superconductor departs from an ideal one. The quasiparticle concentration in superconducting qubits is suggested to exceed the expectation of thermal excitation by orders of magnitudes [1]. Even though the absolute value of quasiparticle concentration is small, decoherence of individual qubits may occur due to nearby quasiparticle states allowing additional channels for single-charge relaxation [2]. For instance, in Josephson junctions, quasiparticles may tunnel from one side of a junction to another [3], leading to qubit decay with a rate proportional to the quasiparticle concentration [4].

Quasiparticle poisoning poses a fundamental challenge in error mitigation when using superconducting qubits. For example, magnetic field is a source of quasiparticle generation [5–7], and even a low field is able to cause an exponential reduction of the qubit parity lifetime [8]. Various methods to control quasiparticle dynamics [9–11], including quasiparticle trap engineering [12–14], were proposed to reduce the poisoning effect to the qubits. Despite that quasiparticle concentration has been characterized on mesoscopic devices via the relaxation rate [10,15–18] or the frequency shift [19,20] measurements, the spatial information of quasiparticle remains unexplored. A direct, nanoscale quantification of the quasiparticle concentration, particularly desired for evaluating the performance of quasiparticle traps, is absent. In this work, we will locally determine quasiparticle concentration by measuring shot noise in a scanning tunneling microscope [Fig. 1(a)].

Shot noise is proportional to the charge of the carriers $q$ and the average current $|I|$, $S = 2q|I|$. Therefore, shot noise offers a direct and sensitive method, via the charge $q$, to investigate minute quasiparticles within a bath of pairs. In the absence of quasiparticles, when the applied bias falls within the superconducting gap energy, the tunneling current into a superconductor is solely coming from Andreev reflections. The Andreev reflection process, i.e., a tunneling electron is reflected as a hole, transfers effectively two electron charge ($q = 2e$), leading to the Andreev current $I_{2e}$. In contrast, direct tunneling into a quasiparticle state with current $I_{1e}$ simply transfers one electron charge ($q = 1e$). In the framework of the tunneling Hamiltonian approach [21], when the transparency of the tunnel junction $\tau$ is small, $\tau \ll 1$, the two current contributions can be formulated as (see Supplementary Materials)

$$I_{ne} = 2ne\tau^n P_n/h, n = 1, 2, \tag{1}$$

with $h$ the Planck constant and the proportional coefficient $P_n$ related to the number of transmitted quasiparticles ($n = 1$) or Cooper pairs ($n = 2$). For a fully gapped superconductor at zero temperature, $I_{1e}$ vanishes below the gap energy because of lack of available quasiparticles to tunnel ($P_1 = 0$). When a finite amount of quasiparticles exist within the gap, the total current is composed of the quasiparticle current and the Andreev current (neglecting higher-order Andreev processes), $I = I_{1e} + I_{2e}$. The shot noise, also consisting of both contributions, can then be expressed as $S = 2q^*|I|$, with $q^*$ as the overall effective charge that can take any value between $1e$ and $2e$ [22, 23]. This $q^*$ value is determined by the quasiparticle concentration $P_1$ and the junction transparency $\tau$. To separate their effects, we define the dimensionless quasiparticle contribution $y = P_1/P_2$. Because $I_{1e}$ and $I_{2e}$ have different dependences in $\tau$, $q^*$ is extremely sensitive to a small portion of quasiparticles, which may not be detectable in the total current $I$. For example, as simulated in Fig. 1(b), for $\tau = 5\times10^{-3}$, quasiparticle contribution $y = 0.02\%$ leads to a reduction of the effective charge from $2.00e$ to $1.95e$, detectable in our present experimental setup.

Here, we employ this extreme sensitivity to quasiparticles by using scanning tunneling noise spectroscopy [Fig. 1(a)] to visualize quasiparticle concentration. As a benchmark, we chose superconducting vortices as the platform to demonstrate our capability of detecting ultralow concentrations of quasiparticles. Superconducting vortices are topological line defects of the superconducting order parameter $\Psi = \Delta e^{i\chi}$. Its phase $\chi$ winds (multiple of) $2\pi$ around the center of a vortex where the amplitude $\Delta$ vanishes [24]. The supercurrent, generated by the gradient of $\chi$ in response to the magnetic field, brings about the emergence of low-energy Bogoliubov quasiparticles [25]. Because the quasiparticle concentration away from vortex cores has a fast, nearly exponential decay in distance [26] and is controllable via an external magnetic field, the vortex lattice system serves as a perfect system for locally generating a minuscule amount of quasiparticles.

We cleave single crystals of 2H-NbSe$_2$ in an ultrahigh vacuum and immediately load it in our scanning tunneling microscope (STM) at a temperature $T = 2.3$ K. We first measure spatially resolved differential conductance, a standard method to image the Abrikosov vortex lattice in NbSe$_2$. Throughout this work, we use a superconducting tip with an energy gap $\Delta_t = 1.3$ meV to achieve an enhanced energy resolution, and the local density of states (DOS) of the NbSe$_2$ sample can be obtained by a standard deconvolution procedure [27,28].

As shown in Fig. 2(a), the differential conductance map at $eV_{bias} = \Delta_t$, which corresponds to the sample DOS at the Fermi level, shows a triangular vortex lattice in an external magnetic field $B = 0.1$ T. In the vortex core a substantial enhancement of the DOS is understood as bound states formed by localized quasiparticles [29], while outside the vortex in the midpoint between two neighboring vortices, the differential conductance spectrum hardly differs from that measured at $B = 0$ T [Fig. 2(d)]. At the bias energy $eV_{bias} = \Delta_t$, the difference in differential conductance between $B = 0$ T and outside the vortex at $B = 0.1$ T is smaller than the error bar of our measurements, which hinders us from detecting residual quasiparticles directly from tunneling conductance.

We now use the noise measurements introduced above, to determine the quasiparticle contributions to the tunneling current. The current noise measured at $B = 0$ T [Fig. 2(e)] follows the $q^*=1e$ line when $|eV_{bias}| > \Delta_t + \Delta_s$ – as expected, because quasiparticles are available outside the gap. When the bias is lowered below the gap energy, $|eV_{bias}| < \Delta_t + \Delta_s$, the noise develops a broadened step transition towards the $q^* = 2e$ curve, indicating that only the Andreev processes contribute to the current and virtually no quasiparticles remain. On the other hand, in a finite field $B = 0.1$ T at the midpoint between two neighboring vortices, the measured shot noise shows a transition departing the $q^* = 1e$ curve for $|eV_{bias}| < \Delta_t + \Delta_s$, but not reaching the $q^*=2e$ curve (see Fig. S5), meaning that a finite fraction of quasiparticle tunneling persists in parallel with the Andreev current. We numerically extract the effective charge $q^*$ for the two cases in Fig. 2(f), and find that the transition within the gap is sharper and reaches a plateau of $2e$ for $B = 0$ T, while for $B = 0.1$ T, it is broader and only plateaus at $1.65e$ outside the vortex at $eV_{bias} = \Delta_t$ (corresponding to the Fermi level of the sample), yielding a portion of 0.14% zero-energy quasiparticles. Note that the differential conductance measurements [Fig. 2(d)] for the two cases look nearly the same, emphasizing the additional information obtained with shot noise spectroscopy.

Next, we quantify quasiparticles in a vortex lattice by spatially resolved shot-noise imaging. Inside the vortex cores, we expect the localized quasiparticles to allow single-electron tunneling, leading to $q^* = 1e$ [28]. Away from the vortex cores, we expect the density of quasiparticles to drastically decrease and $q^*$ to increase above $1e$ because of a major contribution of Andreev reflections to the tunneling current. Therefore, a clear contrast in shot noise is expected between tunneling in and outside the vortex cores. This is exactly what we observe in Fig. 2(b), where we measure shot noise for a constant current, in active feedback, at a fixed bias energy in an area containing 8 vortices. We chose the same field of view as the vortex lattice in Fig. 2(a), which is imaged at the same bias. The resulting spatially resolved effective charge $q^*$ map [Fig. 2(c)] reveals the concentration of quasiparticles, where a darker (lighter) color indicates an effective charge closer to $1e$ ($2e$) and thus more (fewer) quasiparticles. We observe more quasiparticles in the vortex cores, as expected.

Individual vortices in NbSe$_2$ have a peculiar six-fold star shape, as shown in Fig. 3(a). This star shape, which rotates by 30° at higher bias and even reverses its contrast outside the gap [30], is argued to originate from the anisotropy of the superconducting gap and/or Fermi surface [31–37]. As a consistency test, we first extract the spatial dependence of the dynamic junction impedance $R_{dyn}$ from our noise measurements, which is related to the differential conductance measured in active feedback, and thus shows a six-fold star shape in the maps and radial-average plots consistent with the density of state data [Figs. 3(b), 3(d), and 3(e)]. However, our shot noise data around the same vortex in Fig. 3(c) shows an almost isotropic structure. Within our resolution, the radial-average plot of noise [Fig. 3(f)] yields a six-fold anisotropic [$\sin(6\theta)$] term more than one order of magnitude smaller than that of differential conductance [Fig. 3(d)]. We can explain the absence of the six-fold structure in the shot noise map by the insensitivity of the shot noise to quasiparticle tunneling at a higher quasiparticle concentration. Fig. 1(b) illustrates that the effective charge already reaches below $1.05e$ when the quasiparticles are contributing more than 10%. Therefore, when quasiparticle tunneling contribution is high enough, e.g., varying from 10% to 100%, the effective charge only changes by 5%, reaching our present noise resolution.

Finally, we turn to the spatial dependence of quasiparticle concentration between vortices, which is the main focus of this study. The high-resolution shot-noise map around 3 vortices at $B$ = 40 mT shown in Fig. 4 presents the common behavior of all measured vortices. We extract the effective charge $q^*(r)$ at $eV_{bias} = \Delta_t$, where $r$ is the distance along line cuts between two vortex cores in Fig. 4(a). We observe that $q^*(r)$ increases from $1e$ at core centers on both sides ($r = 0$ and $r = d$) to a maximum of $1.78e$, where $d$ is the inter-vortex distance. The maximal $q^*_{max}$ appears at the midpoint between vortices at $r = d/2 = 94$ nm [Fig. 4(b)], where the concentration of quasiparticles is lowest, or the concentration of pairs is the highest. We can describe this spatial dependence with a Ginzburg-Landau model (see Supplementary Materials) in the limit of individual, isolated vortices, if we make a simple assumption that the effective charge $q^*$ is proportional to the pair density. With only one fitting parameter, the coherence length, $\xi = 12$ nm [38], we obtain a good quantitative agreement between our effective charge profile and our Ginzburg-Landau model.

The absence of residual quasiparticles at $B$ = 0 T indicates that the external magnetic field is the cause of quasiparticle states localized in the vortex core and extended between the vortex cores. The question therefore arises how the spatial distribution of quasiparticles depends on magnetic field strength $B$. We carry out shot noise spectroscopy and extract the maximal $q^*_{max} = q^*(r=d/2)$ at the midpoint between vortices at several different field strengths in Fig. 4(c). We observe that $q^*_{max}$ decreases with increasing $B$ with an onset below 40 mT. We note that the estimation of quasiparticles density by the wavefunction overlap of Bogoliubov quasiparticles from the intervortex tunneling (IVT) model [39] does not agree with our observation, especially at the low field limit, as shown by the brown curve in Fig. 4(c). This observation directly confirms the implication that quasiparticles present throughout the vortex state of NbSe$_2$, leading to an onset of increasing thermal conductivity right above the lower critical field $B_{c1}$ = 20 mT [7]. In contrast, our Ginzburg-Landau model fits the magnetic field dependence of $q^*_{max}$ excellently at low field, and it shows that at $B^*$ ~ 1.0 T, $q^*_{max}$ reaches 1$e$ where quasiparticle current dominates the tunneling process in the entire sample.

A ramification of our study is that there is a field $B^*$, much smaller than the upper critical field $B_{c2}$ = 4.0 T at $T$ = 2.3 K of bulk NbSe$_2$ [40], which indicates the emergence of a vortex state with widespread quasiparticles between vortices. Based on this field $B^*$ where $q^*_{max}$ = $q^*(\xi^*=d/2)$ = 1$e$, our model (see Supplementary Materials) shows a limit for inter-vortex distance $\xi^*$ ~ 4$\xi$, below which zero-energy quasiparticle current dominates throughout the whole vortex lattice. However, even when vortices are farther apart than $\xi^*$, quasiparticles already exist between vortices. For instance, at $d$ = 1.4 $\xi^*$ ($B$ = 0.5 $B^*$), $q^*_{max}$ already decreases to 1.2$e$, corresponding to a minimum quasiparticle contribution of one percent.

Our findings are relevant to understanding quasiparticle poisoning in topological qubits using vortex-based Majorana bound states, predicted to exist in iron-based superconductors [41] such as FeTe$_{0.55}$Se$_{0.45}$, where zero-bias peaks in differential conductance have been observed and attributed to Majorana bound states [42,43]. In these reports, the intervortex distance is usually comparable to $\xi^*$ ~ 4$\xi$ = 15 nm for FeTe$_{0.55}$Se$_{0.45}$, although the exact value of $\xi^*$ should be calculated by taking account of properties such as coherence length, Fermi surface, and gap anisotropy. Nevertheless, while previous focus mostly lies on the energy of the lowest-lying states, our local noise imaging uncovers spatial information about quasiparticles that may void the topological protection of putative Majorana qubits: even if Majorana bound states do exist in these vortex cores, our results show an exponential increase of zero-energy quasiparticles generated between vortices. The uncontrolled transfer of zero-energy quasiparticles between these vortices can frequently alter the state of a topologic qubit, which depends on the charge of vortices [44–46].

In summary, we have demonstrated a direct visualization of quasiparticle concentration around superconducting vortices by scanning tunneling shot noise microscopy. Our results show that quasiparticles spread across the vortex lattice when the inter-vortex distance is less than roughly four times the coherence length, which sets a limit for quasiparticle poisoning in vortex-based Majorana qubits. More generally, our noninvasive technique also allows one to locate and quantify of quasiparticle down to atomic scale across common superconducting-qubit devices made of Josephson junctions. By further improving shot-noise resolution below 1 fA$^2$/Hz [47] and implementing our technique in a millikelvin dilution refrigerator, we will be able to measure the background quasiparticle concentration below $10^{-6}$ in thermal equilibrium, and thus allowing direct investigations into the excess quasiparticles in qubits with a concentration of $10^{-9} - 10^{-5}$ [9, 18], whose origin remains to be identified [48].

**Supplementary Material**: See the supplementary material for details on the materials and methods, numerical analysis for the experimental results, and corresponding notes, figures, and additional data.

**Acknowledgments**: We acknowledge U. Vool, C. W. J. Beenakker, C. Bolech, F. van Oppen, V. Geshkenbein, F. Gaggioli, X. Liu, C. Renner, and C. Berthod for valuable discussions. This work was supported by the European Research Council (ERC CoG PairNoise). K.M.B. was supported by the Netherlands Organization for Scientific Research (NWO Veni grant VI.Veni.212.019).

**Author contributions:** J-F.G. and K.M.B. performed the experiments and analyzed the data. All authors contributed to the interpretation of the data and writing of the manuscript. M.P.A. supervised the project.

**Competing interests:** The authors declare that they have no competing interests.

**Data and materials availability:** All data in the main text and the supplementary materials will be available on Zenodo at https://doi.org/10.5281/zenodo.10689647.

**Figures**

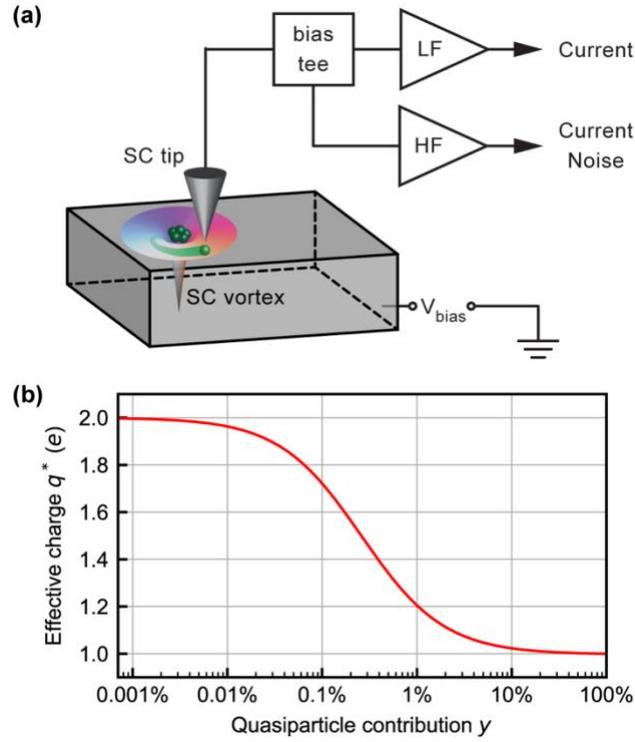

FIG. 1. Visualizing quasiparticles by scanning tunneling noise spectroscopy. (a) Schematic illustration of the scanning tunneling noise microscope setup. A bias voltage $V_{bias}$ is applied between the superconducting (SC) tip and sample. A SC vortex is shown in the sample: the order parameter has a winding phase (color wheel) and a decreasing amplitude (height) inside the vortex core. Green balls illustrate quasiparticles. HF and LF stand for high- and low-frequency amplifiers, respectively. (b) Simulation of effective charge as a function of quasiparticle contribution at a fixed junction transparency $\tau = 5\times10^{-3}$ at $T = 2.3$ K. The effective charge $q^*$ is extracted numerically by solving $S=2q^*|I|\coth(q^*V/2k_BT)$.

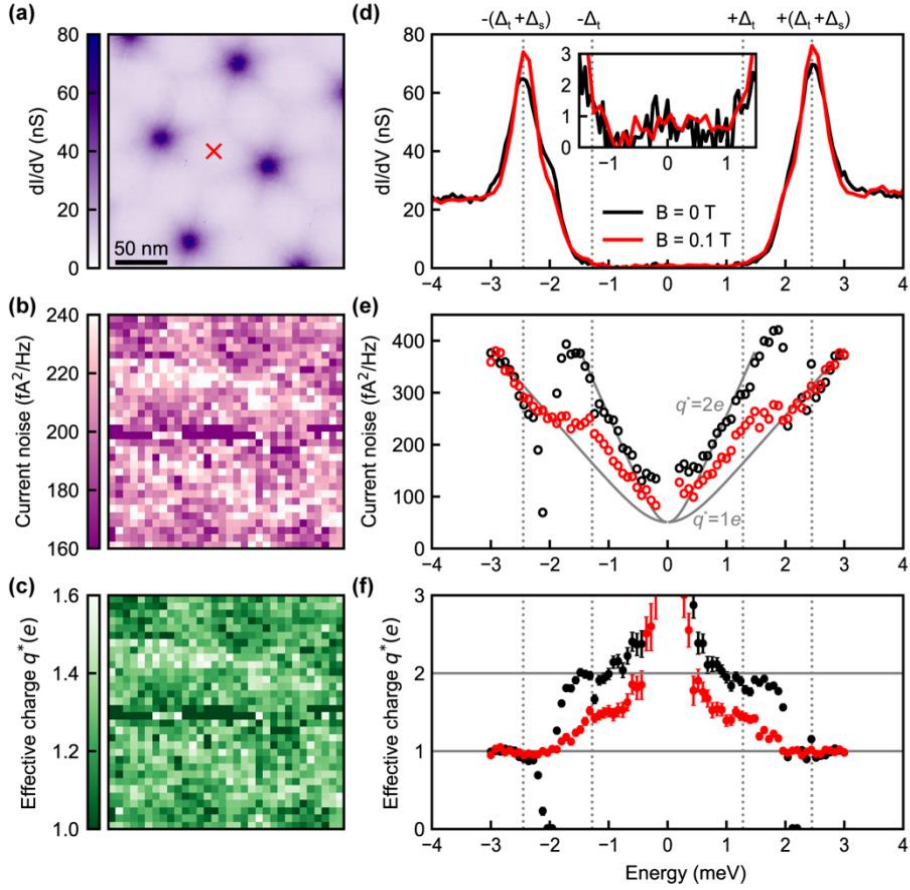

FIG. 2. Differential conductance and noise spectroscopic imaging on vortices of NbSe$_2$. (a) Differential conductance image measured at $eV_{bias} = \Delta_t$ and $B = 0.1$ T showing a lattice of vortices. (b) Spatially resolved current noise measured at $eV_{bias} = \Delta_t$ in the same field of view as (a). (c) Effective charge image extracted from (b). (d) Differential conductance spectra taken in zero field (black, at a random position) and $B = 0.1$ T (red) at the red cross marked in (a). The inset shows a zoom-in view of the spectra inside the gap. Noise spectra (e) and the extracted effective charge (f) at the same locations as (d). The gray curves in (e) are the expected junction noise with an effective charge $q^*$ of $1e$ and $2e$ at $T = 2.3$ K. The error bars are determined by the fluctuation of the current noise in time, yielding a standard deviation of 9.25 fA$^2$/Hz. The dotted lines in (d–f) indicate peak energies $\pm(\Delta_s + \Delta_t)$ and gap energy of the tip $\pm\Delta_t$. Setup conditions: (a) and (d), $V_{set} = 5$ mV, $I_{set} = 200$ pA; (b), $V_{set} = 1.3$ mV, $I_{set} = 520$ pA, $R_J = 2.5$ MOhm; (e), $R_J = 2.5$ MOhm.

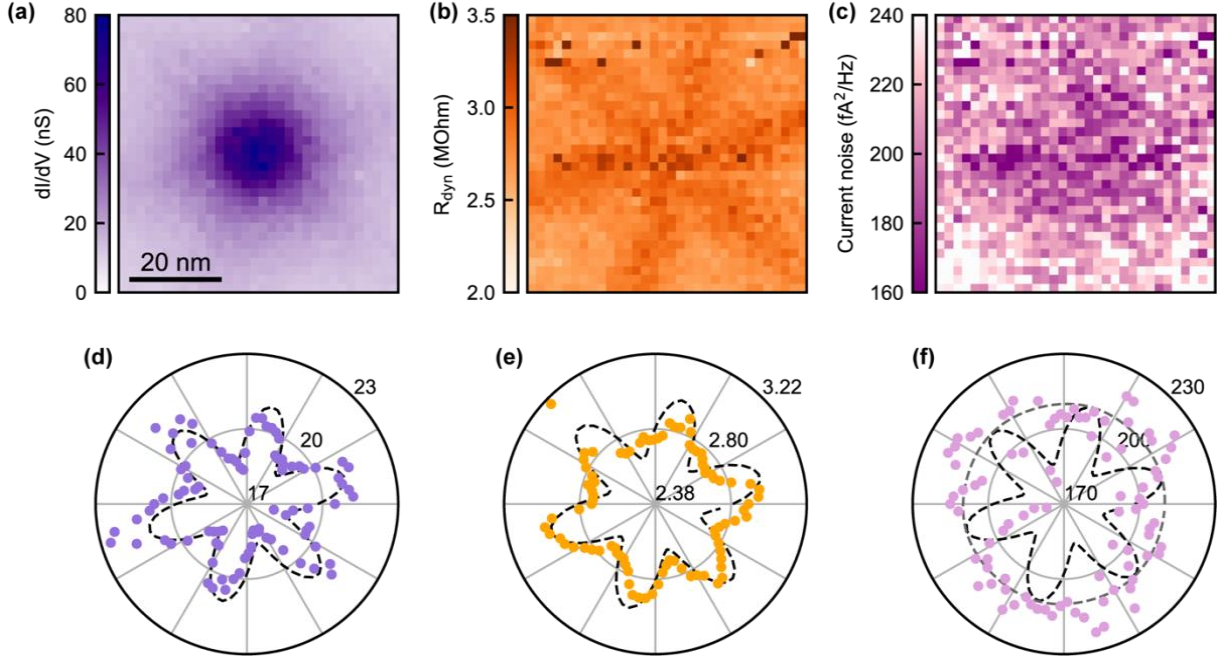

FIG. 3. Core structure of an individual vortex. (a) Differential conductance image measured at $eV_{bias} = -\Delta_t$ and $B = 0.1$ T. Dynamic resistance (b) and current noise (c) imaged at $eV_{bias} = -\Delta_t$ in the same field of view as (a). (d–f) Radial average of (a–c), respectively, in the radial range of 19.3 nm – 28.0 nm from the vortex core (see Fig. S4 for details). The black dashed, six-fold star curve $A[1+0.06\sin(6\theta)]$ serves as a guide to the eye for spatial anisotropy. Here $A$ is the azimuthal-averaged amplitude of 19.8 nS for (d), 2.81 MOhm for (e), and 198 fA2/Hz for (f). The gray dashed line in (f) shows a weaker anisotropy $A[1+0.001\sin(6\theta)]$ for comparison. Setup conditions: (a), $V_{set} = -5$ mV, $I_{set} = 200$ pA; (b) and (c), $V_{set} = -1.3$ mV, $I_{set} = 520$ pA, $R_J = 2.5$ MOhm.

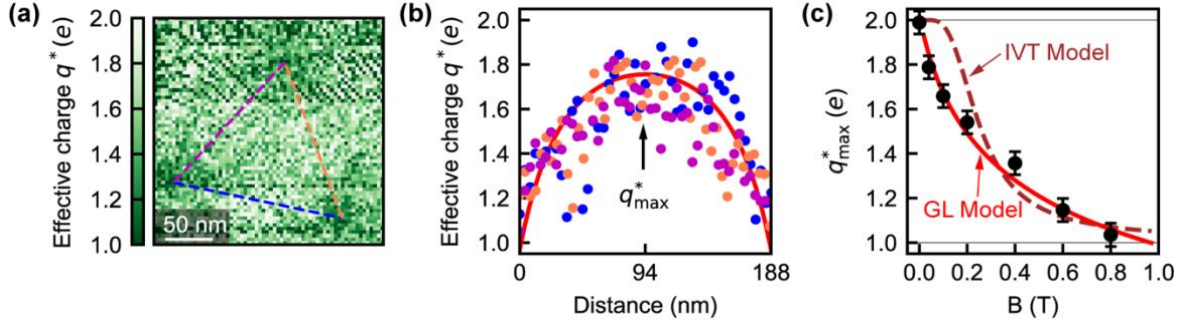

FIG. 4. Imaging quasiparticle concentration around three vortices. (a) Effective charge image measured at $eV_{bias} = \Delta_t$ and $B = 0.04$ T. Setup conditions: $V_{set} = 1.3$ mV, $I_{set} = 520$ pA. (b) Line profiles of effective charge along three linecuts between centers of vortex cores in (a). (c) Magnetic field dependence of the maximal effective charge $q^*_{max}$. The error bars are determined by the standard deviation of the extracted $q^*$ in the energy ranges ($\Delta_t \pm 0.1$ meV) and -($\Delta_t \pm 0.1$ meV) in Fig. S2. The red lines in (b) and (c) are the expected effective charge from GL model fit, Eq. (S4), with $\xi = 12 \pm 2$ nm. The brown dashed curve is the inter-vortex tunneling model following Eq. (S1), with $B_{c2} = 4.0$ T.